# Advancement of Deep Learning in Pneumonia/Covid-19 Classification and Localization: A Qualitative and Quantitative Analysis

**Aakash Shah**
*Department of Computer Science & Engineering, Institute of Technology, Nirma University, Ahmedabad, Gujarat, INDIA*

**Manan Shah**
*Department of Chemical Engineering, School of Technology, Pandit Deendayal Petroleum University, Gandhinagar, Gujarat, INDIA*

**Abstract**

Around 450 million people are affected by pneumonia every year which results in 2.5 million deaths. Covid-19 has also affected 181 million people which leads to 3.92 million casualties. The chances of death in both of these diseases can be significantly reduced if they are diagnosed early. However, the current methods of diagnosing pneumonia (complaints + chest X-ray) and covid-19 (RT-PCR) require the presence of expert radiologists and time, respectively. With the help of Deep Learning models, pneumonia and covid-19 can be detected instantly from chest X-rays or CT scans. This way, the process of diagnosing pneumonia/covid-19 can become faster and more widespread. In this paper, we aim to elicit, explain, and evaluate, qualitatively and quantitatively, all advancements in deep learning methods aimed at detecting community-acquired pneumonia (CAP), viral pneumonia, and covid-19 from images of chest X-rays and CT scans. Being a systematic review, the focus of this paper lies in explaining various deep learning model architectures which have either been modified or created from scratch for the task at hand. For each model, this paper answers the question of why the model is designed the way it is, the challenges that a particular model overcomes, and the tradeoffs that come with modifying a model to the required specifications. A grouped quantitative analysis of all models described in the paper is also provided to quantify the effectiveness of different models with a similar goal. Some tradeoffs cannot be quantified, and hence they are mentioned explicitly in the qualitative analysis, which is done throughout the paper. By compiling and analyzing a large quantum of research details in one place with all the datasets, model architectures, and results, we aim to provide a one-stop solution to beginners and current researchers interested in this field.

## 1. Introduction

Pneumonia is an infectious respiratory disease responsible for significant morbidity all over the world. It causes a lower respiratory tract infection, leading to inflammation in the lungs' air sacs known as alveoli. The infected alveoli are filled with fluid which makes breathing difficult. Pneumonia, a contagious disease, is classified into two main types (Hospital-Acquired Pneumonia & Community-Acquired Pneumonia) based on where it is acquired. The majority of pneumonia cases fall under the category of Community-Acquired Pneumonia (all cases of pneumonia that are not acquired from the hospital), commonly referred to as CAP. If CAP is diagnosed early, the chances of 100 percent recovery are high, with little chances of re-infection. For a complete diagnosis of pneumonia, a combination of clinical awareness, specific microbiological tests, and radiographical studies are necessary. However, plain chest radiography alone can rapidly demonstrate the presence of pulmonary abnormalities in most cases 1. Unfortunately, pneumonia is only one of many pulmonary abnormalities, and hence, radiographical findings often fail to lead to a definitive diagnosis of pneumonia. Consequently, the distinction of pneumonia from other pulmonary diseases cannot be made with certainty on radiological grounds with current technology.

One of the significant problems of radiographical findings is that the distinction of pneumonia from other pulmonary diseases cannot be made with certainty on radiological grounds alone. Moreover, this is not the only problem with the current procedure of pneumonia diagnosis. A considerable number of medical images are produced in hospitals and medical centers daily. Consequently, radiologists are inundated with a large number of images that they have to analyze manually. In these cases, tried and tested deep learning algorithms might be helpful in assisting doctors by marking the part of the lungs where pneumonia/covid-19 is present.

Many automated technologies related to medical imaging have shown promising results over the past few years, but deep learning has quickly gained prominence amongst them. Researchers have extensively exploited deep learning methods for detecting diseases in various body parts such as the eye, brain [2], [3], and skin [4], [5]. In some medical imaging cases, it was shown that the classification performance of a DL model was better than that of medical specialists [6]. Since the proposal of AlexNet [7] in 2012, deep learning models have improved significantly in image classification tasks. Recent architectures like ResNet and variations of ResNet have also provided a solid base for accurate object detection and localization. While single-shot detectors like Yolo [8] and RetinaNet [9] provide speedy detections useful in real-time, Generative Adversarial Networks [10] have played an essential role in unsupervised learning and domain adaption whenever training images have been scarce. Hence, automated deep learning solutions can solve both problems mentioned above. Deep learning models for pneumonia classification and detection can automatically learn complex features from radiographs that may not be visible to the naked eye. This was proved in 2017 when (Rajpurkar et al.) [6] proposed CheXNet, a deep learning model, which achieved better results than radiologists on pneumonia detection and other pulmonary disease detection tasks.

The fact that deep learning models succeeded, not only in the task of pneumonia detection but also in other pulmonary abnormality detection tasks, was leveraged by many other researchers to detect other anomalies from the same models or training data. This use case could prove useful, especially in recent situations (in 2021) like the outbreak of Covid-19 because of the following reasons. Even though real-time polymerase chain reaction (RT-PCR) is the accepted standard in the diagnosis of covid-19, its sensitivity and specificity are not optimal [11]. Other than that, many countries or regions cannot conduct sufficient RT-PCR testing for thousands of subjects in a small span of time because of the lack of people who can perform these tests. In these cases, deep learning algorithms might help if the country has enough imaging machines but fewer people who can perform the test. RT-PCR testing may also be delayed in cases of newly evolved coronavirus because detection of a newly evolved virus requires the extraction of the new DNA sequence [11]. In contrast, deep learning models with anomaly detection capabilities can detect the clustering effect of viral pneumonia occurrences like MERS (W. Li, 2004.), SARS (Azhar, 2014.), and COVID-19 as proved by [11]. Thus, deep learning models provide a vital technique that might help in diagnosing pneumonia better and faster.

In this paper, we aim to elicit, explain, and evaluate, qualitatively and quantitatively, all advancements in deep learning methods aimed at detecting bacterial or viral pneumonia from radiographical images. Since Chest X-rays and CT scans are the most common radiographical tools doctors use today, we have covered deep learning methods that use Chest X-rays, CT scans, or both as input images. Because the quantitative results of these models depend on the datasets used, we group these models according to datasets in order to perform a fair and uniform quantitative analysis. While standard datasets are available for bacterial/viral pneumonia detection tasks, the same is not applicable for covid-19 datasets due to the disease's novelty (in 2021). However, the models that leverage these datasets have been grouped by the amount and quality of images used for training and testing. This being said, it is not uncommon to find deep learning models that fail to perform well in the real world after being trained on datasets with specific sources. The poor performance in the real world is mainly because of the dataset shift between training images and the images used in other hospitals. A significant amount of variability in individual hospital images also accounts for the poor performance of these models. To address this problem, we also evaluate and compare the features learned by various models to predict how well they would perform in the real world. The reason for comprehensively compiling all significant research in deep learning for pneumonia detection is to compare different models used in each scenario and identify the best deep learning architectures for each of those scenarios. Although similar work was performed by (Y. Li et al.)[14], we provide a significantly more comprehensive overview of models by including research with CT scans, localization tasks, and covid-19 classification.

## 2. Methodology

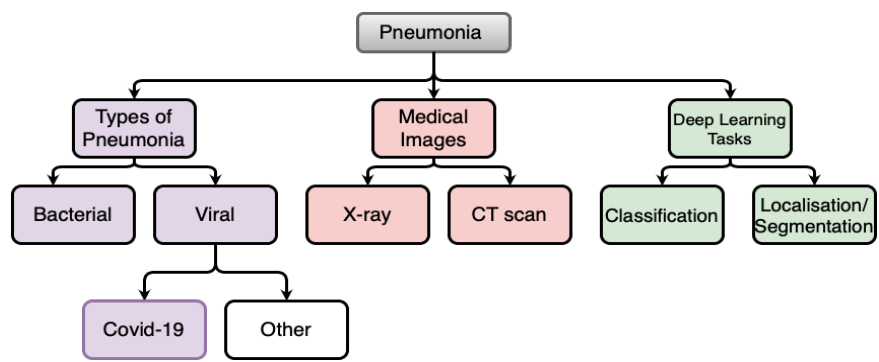

*Figure 1*

This review is based upon the qualitative and quantitative analysis of studies in the field of pneumonia/covid-19 detection via Chest X-rays and CT-Scans. The method for collecting relevant papers for this study was as follows. Platforms like Elsevier, Google Scholar, IEEE Xplore, and Springer were searched with the keywords: "pneumonia detection with deep learning", "covid-19 detection with deep learning", "pneumonia localization with deep learning", "covid-19 localization with deep learning", "pneumonia detection with Chest X-rays", "pneumonia localization with chest X-rays", "covid-19 detection with chest X-rays", "covid-19 localization with chest X-rays". Papers were excluded from the study as follows. All papers not related to deep learning, pneumonia, or covid-19 were excluded. After the first exclusion process, all remaining papers were included in the final review according to the following criteria. As the main focus of this review is on the generalizability of models, all studies which made an explicit effort to make their model generalizable were included. Different studies used various metrics for accuracy, so there was no hard limit of accuracy (performance in general) for a paper to be included in this study. After that, studies were included with the goal of covering as much breadth in deep learning methods as possible. This was done because different deep learning methods often solve different problems (improper images, training data shortage, insufficient training data variety). Furthermore, if a similar method was followed by more than one paper, then the most generalizable and the paper with the best performance was chosen.

On the medical front, pneumonia is mainly divided into two types. 1) Bacterial Pneumonia and 2) Viral Pneumonia. While bacterial pneumonia does not have any sub-categories worth discussing here, viral pneumonia is often sub-categorized according to the virus responsible for causing viral pneumonia. The most recent example of viral pneumonia and of concern to us is covid-19. Owing to these types and sub-types, researchers broadly classify input images into 1) pneumonia/no-pneumonia, 2) bacterial pneumonia/viral pneumonia/no-pneumonia and 3) Covid-19/all other pneumonia/no-pneumonia. While most research papers fall into one of these three categories, some models do not consider no-pneumonia.

Radiologists use either 1) Chest X-rays or 2) CT scans for diagnosing a patient. Both of these modes have their pros and cons. While X-ray machines are portable and enable faster diagnosis, CT scans provide finer detail of the lungs that may be more difficult to see in a plain X-ray. Similarly, some deep learning models use X-rays as input images, while others use CT scans. This paper gives equal weightage to both models mentioned above but discusses them separately in sections III and IV, respectively.

Other than classification, a significant task taken up by some DL models is that of detecting and localizing the region where pneumonia is present in the lungs. It is worth noting that some classification models also perform grad-cam analysis to analyze which features are being used to perform classification. These models, even after localizing features, are not considered localization/segmentation models. Localization/Segmentation models provide bounding boxes/semantic segmentation in input images around the part of the chest affected by pneumonia. We will include these models in our discussion too. However, their comparison shall only be made with other localization models.

| Dataset | Images | Classes | Bounding Boxes |
|---|---|---|---|
| NIH Chest X-rays | 1,12,120 | 14 | 985 |
| RSNA Chest X-rays | 26,684 | 3 | 9555 |
| Kaggle Chest X-rays | 5,856 | 3 | 0 |
| CheXpert | 2,24,316 | 14 | 0 |
| MIMIC-CXR | 3,71,920 | 14 | 0 |

*Table 1*

Datasets play one of the most prominent roles in the success or failure of deep learning models. The details of the three most frequently used datasets are as follows. The **NIH** dataset consists of 15 classes, out of which one is pneumonia, one is no pulmonary disease, and the remaining 13 are other pulmonary diseases. It is worth noting that "other pulmonary diseases" may have any number of classes ranging from 0 to 13. This way, if it has 0 classes, the classification task simplifies to pneumonia/no-pneumonia (1 sigmoid neuron or two softmax neurons in the output layer). On the other hand, if it has 13 classes, the model will classify a chest X-ray into pneumonia, no-pneumonia, or any one of the 13 pulmonary diseases (15 softmax neurons in the output layer). The classes of the **RSNA** dataset are Normal, Lung opacity, and No lung opacity-not normal, which can be explained as no pneumonia, pneumonia with visible lung opacity, and some pulmonary disease without visible damage to the lungs. Lastly, the classes of the Kaggle dataset are divided as Normal, Bacterial-Pneumonia & Viral-Pneumonia that need no further explanation.

**3.1 Detection of Pneumonia and its Classification amongst other Pulmonary Diseases**

Rajpurkar et al. [6] developed a DL model that could achieve radiologist-level accuracy on pneumonia detection from chest X-rays. They used the NIH dataset, which consists of 112,120 chest X-ray images from 30,805 patients. This dataset was first presented and used by [15] for the same task. However, the model given by [6] was the first one that attained radiologist-level accuracy, and it also served as a base for many future models. Firstly, the entire dataset is split into training and test sets such that no patients are repeated in the respective sets. The images are converted to size 224 X 224 and normalized by the ImageNet [16] training dataset metrics. For training, these images are fed into the CheXNet model that uses a 121 layered Dense CNN known as DenseNet [17]. DenseNet improves information flow and backpropagation through the network, which makes the optimization process easier. Hence, the entire model was used as it is except for the output/classification layer. This layer was replaced by a single sigmoid neuron because the classification task was pneumonia/no-pneumonia. Because the NIH dataset consists of 15 classes, the classes pneumonia and no-pneumonia (14 classes including other pulmonary diseases) were highly imbalanced. To get rid of this problem, a weighted loss function is used while training the model. Finally, the model achieved an F1 score of 0.435 and an AUROC of 0.76 when tested with 420 images. The dataset was randomly split into training (28744 patients, 98637 images), validation (1672 patients, 6351 images), and test (389 patients, 420 images). There was no patient overlap between the sets.

Zech et al. [18] demonstrated that deep learning pneumonia classifiers trained on two different hospital systems predicted results by learning the origin of those hospitals instead of learning relevant features that cause pneumonia. To address this problem, Janizek et al. [19] developed an adversarial training-based approach. They found that the occurrence of pneumonia in PA (posterior-anterior) chest X-rays was twice as much as that of pneumonia in AP (anterior-posterior) images. (PA images are ones in which X-rays enter from the back of the body while AP is vice-versa). They also found out that pneumonia detection classifiers as in [6] learned to distinguish between the two views (AP and PA) and leveraged that information to classify pneumonia. Their approach was different from standard adversarial approaches, where the classifier learns domain-invariant features. In their case, the classifier could not learn domain-invariant features because they had no images from the target domain. In their adversarial approach, Janizek et al. [19] tried to train a classifier in which the final output score of the classifier would be invariant of the view (AP or PA). While the training and architecture for their classifier were the same as that of [6], they also added and trained an adversary network. This adversary network took the output score of the classifier as input and outputted a prediction of the view. The adversary network is a standard 3 layered feed-forward-network of 32 neurons, each with ReLU activations. The classifiers' objective was to predict output scores such that the adversary could not predict the view of the input image from the output score. In contrast, the adversarial network's objective was to predict the output score's view (AP or PA). Both the classifier and the adversary network were trained alternatively for optimizing their respective objectives. To test their approach, Janizek et al. [19] tested their model on the CheXpert dataset (source domain) and MIT's MIMIC-CXR dataset (target domain). While the standard model (without the adversary network) achieved an AUROC of 0.79 on the source domain, it could only achieve an AUROC of 0.703 on the target domain. Alternatively, the adversarially trained model achieved almost similar AUROC's of 0.747 and 0.739 on the source and target domains.

In April 2020, Lu et al. [20] presented the MUXConv, a CNN layer specially designed to increase the flow of information by multiplexing channels and spatial input through the network. They also presented a multi-objective algorithm to automatically optimize hyperparameters while training. Though the MUXConv was not specially designed for pneumonia classification, it could achieve an AUROC of 84.1% on the same dataset used by [6] while using 3x fewer parameters, being 14x more efficient than DenseNet-121 and without any manual hyperparameter optimizations. This result shows the scope of improvement in the accuracy of pneumonia detection through better Deep Learning architectures alone, i.e., without considering any medical knowledge. In September 2020, the same team presented the NSGANetV1, another multi-objective evolutionary algorithm. NSGANetV1 learns the designs of various architectures through the recombination and generation of multiple architectural components. NSGANetV1 makes its efficiency better by exploiting various patterns used in successful architectures by estimating their distributions with the help of a Bayesian model. Though made for general-purpose image classification, this model achieved an AUROC of 84.6% on the NIH dataset without modifications or hyperparameter tuning. Moreover, the class activation map of NSGANetV1 showed that the model learns relevant features, which can also be used to pinpoint the region where pneumonia is present.

Using architectures like DenseNet-121 in the pneumonia detection task is possible because of large datasets like NIH or CheXpert. If such architectures are used with smaller datasets like that of Kaggle, there is a considerable chance of overfitting. Li et al. [21] presented the PNet, an efficient yet effective architecture for pneumonia detection using a significantly smaller number of images. They collected their own dataset from Shenzhen No.2 People's Hospital, consisting of 6339 X-rays labeled pneumonia and 4445 X-rays labeled normal. The architecture of PNet is straightforward, consisting of only 5 convolution blocks, each followed by a max-pooling layer. This small architecture allows PNet to be 25 times as efficient as AlexNet and approximately 50 times as efficient as VGG 16. detection task with an accuracy of 92.79 % and an F1 score of 0.93. Even though PNet has a smaller number of parameters, it outperforms both the AlexNet and VGG 16 in the pneumonia are many customized architectures like PNet, which also get equivalent accuracy. However, only PNet was included in our research because of its excellent results on feature analysis. While analyzing the features of all models, it was found that VGG 16 focuses on the entire lung region instead of focusing on the pneumonia-affected region and AlexNet wanders off to the wrong regions. On the other hand, PNet focuses on only those features which correspond to the pneumonia-affected region in most cases. Hence, PNet is not only good at detecting pneumonia, but it can also help doctors by highlighting the pneumonia-affected area. The detailed results were TP/FP/TN/FN: 617/86/360/19 with a sensitivity of 0.9701 and specificity of 0.8072.

Yumin et al. [22] presented a network architecture that achieved high classification accuracy in pneumonia detection. They used an Improved Quantum Neural Network and trained this model on the Kaggle Chest X-ray dataset containing 5232 training images. This model was tested using 624 separate images in the test set and achieved an accuracy of 96.07 %. They also trained AlexNet, ResNet, and InceptionV3 on the same data, giving 85.3%, 86.38%, and 95.53% accuracy. Although the authors do not conduct a feature analysis in their paper, chances are few that a Quantum Neural Network would give such high accuracy while learning wrong or irrelevant features. The dataset that these authors used was published by the University of California, San Diego. The sensitivity and specificity were 0.9756 and 0.9460, respectively.

Diving deeper into pneumonia detection with small datasets, most intuitively, we come across a solution based on Generative Adversarial Networks. Khalifa et al. [23] used a GAN with various deep learning models to generate more images and use those images to train the deep learning models. They took only 10% images from the Kaggle Chest X-ray dataset and generated the remaining 90% with the GAN for training purposes. These images were then used for training by AlexNet, SqueezeNet, GoogleNet, and ResNet with 8, 18, 12, and 18 layers, respectively. ResNet performed best with a testing accuracy of 99.0% and a recall of 0.9897. The catch, however, is that they used 624 images to train the GAN, which is the same number of images provided in the testing dataset. While the authors have mentioned that three separate trials were conducted with a different 10% of the dataset, using test images in even one of the four trials would drastically change the average accuracy. Nonetheless, the idea of using GAN's to generate new data can certainly be applied when there is a dearth of training images.

Dey et al. [24] developed a model with an Ensemble Feature Scheme (EFS) for pneumonia detection. Their EFS combines hand-crafted features and automatically extracted features from a deep learning model to classify an image into pneumonia or normal. Extraction of hand-crafted features is again completed by combining Continuous Wavelet Transform (CWT), Discrete Wavelet Transform (DWT), and GLCM (Grey Level Co-occurrence Matrix). The deep learning features are extracted using the standard VGG-19 architecture. The combined hand-crafted features are then concatenated with features extracted using VGG-19 through PCA and serial feature concatenation. After concatenation, these features are given as an input to a random forest classifier for final classification. This model was trained using 5500 images from the NIH dataset and achieved 97% accuracy when tested against 1650 separate images from the NIH dataset. Like other models mentioned in this paper, the feature activations of this model also point to relevant regions in the lung where pneumonia is present. The detailed metrics were TPR/FPR/TNR/FNR: 0.9756/0.0244/0.9808/0.0192 with a sensitivity of 0.9807 and specificity of 0.9757

| Author | Model | Dataset | AUROC | Accuracy |
|---|---|---|---|---|
| Rajpurkar et al. [6] | CheXNet (DenseNet121) | NIH | 0.760 | NA |
| Janizek et al. [25] | CheXNet (DenseNet + Adversarial) | NIH + MIMIC | 0.747 | NA |
| Lu et al. [20] | MUXConv (Multiplexed Convolutions) | NIH | 0.841 | NA |
| Lu et al. [20] | NSGANetV1 | NIH | 0.846 | NA |
| Li et al. [21] | P-Net (Customized CNN) | Custom (10,784) | NA | 92.79% |
| Yumin et al. [22] | Quantum Neural Network | Kaggle | NA | 96.07% |
| Khalifa et al. [23] | GAN (semi-supervised) | Kaggle (624) | NA | 99.00% |
| Dey et al. [24] | EFS (CWT + DWT + GLCM) | NIH (5550) | NA | 97.00% |

*Table 2*

**3.2 Detection of Covid-19 and Classification of Viral Pneumonia from Bacterial Pneumonia**

Capturing a Chest-Xray is one of the primary methods of screening the occurrence of Covid-19. However, there is a general dearth of doctors even at places where equipment to capture such X-rays is available. To tackle this problem, a lot of research has been done to detect Covid-19 from Chest-X-rays automatically. Cases of Covid-19 emerged in the entire world in 2019, but a lot of research in pneumonia detection from Chest-X-rays had already been done before. Hence, much research on the detection of Covid-19 from Chest-X-rays is built upon the base provided by previous research into pneumonia detection. Due to the novelty of Covid-19 (in 2020-21), no standardized databases are available, and almost every research work uses a different database. Hence, the details of all databases and comments on their quality are given while explaining the research work rather than giving an overview of all databases beforehand.

Haghanifar et al. [26] made a hierarchical DL model for detecting Covid-19. In the first level, images of Chest X-rays are classified into normal and pneumonia. In the second level, images classified as pneumonia are further classified into Covid Positive (CP) or Community-acquired Pneumonia (CAP). The dataset used by the authors contains 780 Covid-19 positive X-rays, 4600 X-rays having CAP, and 5000 Normal X-rays. The approach taken by [26] was very similar to that of [6]. The key difference was that [26] first segmented the lungs from Chest-Xray, and then they only used the part surrounding those lungs for classification. This approach, to a significant extent, solved the issue of "Learning the wrong features to reach the right answer" because then, the model was forced to learn only from the lung region rather than learning from the entire X-ray, which usually contains a lot of regions other than the lungs. U-Net was used for segmentation of the lung region, and then they performed dilation on the segmented lungs to cover some lung areas that the U-Net did not segment. After segmentation, they cropped the Chest-Xray image such that only the segmented area was covered. This cropped image was then fed into the DenseNet-121 model given by [6]. This model achieved an accuracy of 81.04% and f-scores of 0.85 and 0.76 for Covid Positive and Community-acquired Pneumonia classes, respectively. While the accuracy of this model is 0.4% less than that of CheXNet [6], it is more robust than CheXNet on unseen data because of the cropped images. The precision and recall for (Normal/Pneumonia/Covid-19) were P: (0.8251/0.9340/0.9420) and R: (0.9516/0.7797/0.9420), respectively.

While on the topic of lung segmentation, we cover another research work by [27], which uses lung segmentation to classify a chest X-ray into bacterial pneumonia or viral pneumonia. The dataset used by them consists of 241 X-ray images where lungs have been separated manually. The rest of the dataset consists of 4513 pediatric chest X-ray images, out of which 2665 are Bacterial pneumonia and 1848 are Viral pneumonia. The entire model is divided into three parts. The first part is where the lung region is segmented from the chest X-ray by an eight-layer FCN [28]. The FCN model was trained using the 241 segmented images from the JSRT dataset and used pre-trained weights from the Pascal VOC [29] segmentation dataset. The second part consists of feature extraction, where features are extracted using three different methods. The first method uses a DCNN; the second method uses a mixture of GLCM based texture features and HOG-based shape features, while the third method uses HAAR wavelet texture features. The third part of the model uses a simple SVM classifier to classify a given image into bacterial pneumonia or viral pneumonia. This particular approach achieved an accuracy of 76.92% with an AUC of 82.34%. At this point, it is imperative to reiterate that metrics like accuracy, F-scores, and AUC should not be the only parameters to judge the performance of a DL Model. In fact, in most cases, perfect or close to perfect metrics suggest the opposite of sound because in most cases, the underlying model is overfitted, not because of the complexity of the model or the lack of data, but because of learning irrelevant features that are specific to the source of train data. The model achieved a sensitivity of 0.5567 and specificity of 0.9267.

Covid-19 is a type of viral pneumonia, but it is not the only type of viral pneumonia. Several different respiratory diseases such as MERS and SARS fall into the category of viral pneumonia. Moreover, the occurrence of clusters of viral pneumonia cases over a short period can be a signal of an upcoming outbreak or a pandemic. Keeping this in mind, Zhang et al. [11] developed a Confidence Aware Anomaly Detection (CAAD) model to detect the occurrence of viral pneumonia from chest X-rays. To train their model, they used two in-house datasets named X-Viral and X-Covid. The X-Viral dataset contains 5,977 viral pneumonia images, 18,619 non-viral pneumonia images and 18,774 normal images. The X-Covid dataset contains 106 Covid Positive images and 107 normal images. They also used the Open Covid dataset containing 493 Covid Positive images. The CAAD model has 3 main parts. A feature extractor, an anomaly detector, and a confidence predictor. Before we go any further, it is essential to clarify that the "anomaly" we are trying to predict is viral pneumonia, and all other classes (pneumonia and normal) are considered normal. Moving back to the model, after passing an image to the feature extractor, the features are passed simultaneously into the anomaly detector and the confidence predictor. If the anomaly detector predicts the image as an anomaly or the confidence predictor predicts our model's confidence below a particular threshold, the image is considered an anomaly, i.e., viral pneumonia. The feature extractor is made up of EfficientNet B0 [30]. The authors designed the anomaly predictor and the confidence detector, and they are not as common as other ones mentioned in this review, so they deserve an explanation. However, the explanation is too involved and out of the scope of this review, so readers are requested to read the original paper for an explanation of those modules. Coming to the results of this approach, it achieved 80.33% accuracy on the X-viral dataset with training and 78.57% accuracy on the X-Covid and Open-Covid datasets combined without any training. This shows us that the model could categorize Covid-19 cases as viral pneumonia without any specific training on Covid-19 images, which shows that this model can be useful in predicting upcoming cases and different mutations of viral pneumonia. The sensitivity and specificity on various datasets for viral and normal classes were: (X-Viral: 85.88/79.44), (X-Covid: 71.70/73.83), (Open-Covid: 100/100), (X-Covid + Open-Covid: (77.13/78.97)).

Another instance of a region-based discriminator for Covid-19 was given by [31] in August 2021. They used the Covid-CXR dataset consisting of 204 Covid Positive X-rays and the RSNA pneumonia detection dataset for 2004 CAP and 1314 Normal chest X-rays to train their model. The authors proposed a Discrimination-DL and a Localization-DL, but their approach was completely different. They divided all chest X-ray images into superpixels first, and then they ran a proposal of lung (POL) regressor over those superpixels. This approach is very similar to that of YOLO [8], with a critical difference that only the outer boundaries of all superpixels inside the POL proposed rectangles are used to extract two lungs. After both lung regions are extracted, they are passed into the Discrimination-DL, which comprises a ResNet and a feature pyramid network over the ResNet to rebuild the image after feature extraction. Focal loss is then measured against the rebuilt image, and the original lung region is passed into the discrimination-DL. This method helps the Discriminator-DL in learning optimal features. If the Discriminator-DL classifies the image into Covid Positive, both the softmax score and original image are passed into the Localization-DL. The Localization-DL only gives the 1 out of 3 results, i.e., it classifies the Covid-19 as either present in the left lung or the right lung or both lungs. The name Localization-DL is might thus seem to be misleading because it is more of a classifier. Nevertheless, the Localization-DL uses a residual attention mechanism to determine the occurrence of Covid-19 in both lungs. The residual attention mechanism looks at the features extracted by the feature extractor to determine where the attention of the classifier lies. For a deeper analysis of the residual attention mechanism, the reader is referred to the original paper [32]. Coming to the accuracy of this model, it achieves 99%, 90%, and 93% accuracy on Covid Positive, Community-acquired Pneumonia, and Normal classes, respectively.

Arias-Londono et al. [33] presented a thoughtful evaluation approach for DL networks that detect Covid-19. Not only that, but they also compiled the most extensive known dataset of 8573 unique Covid-19 chest X-rays. The entire dataset consisted of 49000 Normal, 2400 Community-acquired Pneumonia, and 8573 Covid-19 Positive images. They used the same DL model used in Covid-Net [34] and ran 3 different experiments on this dataset and model. The first experiment used raw images as an input, with the only pre-processing being histogram equalization. In the second experiment, they used U-Net to segment the lung region and cropped the image so that only the region encompassing the two lung regions remained. In the third experiment, the same segmentation approach was used, but this time they only kept the segmented lung part while the remaining region was filled with a black mask. Upon Grad-Cam analysis, it was found that only experiment 3 learned relevant features even if the accuracy was lower than that of the other two experiments. They also showed that the accuracies of the AP X-ray projection were significantly higher than that of the PA projection. The showings of this research take us to an important point worth noticing. As shown below, metrics like accuracy and F-scores can be bolstered if the DL model is not extracting the right features. However, models made in such a manner may be poor at generalizing to new data from a new source. Hence, Grad-Cam analysis is crucial to determine whether a given model will be able to perform well in the real world, and one should not judge a model solely based on its metrics, especially if the train/test data is less or if the train/test data belong to the same source.

Before we continue with our quest for the best DL models for covid-19 detection and classification of viral pneumonia from bacteria pneumonia, we should make a note. The constructions of all models discussed above show an explicit effort to make the model perform well in the real world. These efforts are shown in the form of Grad-Cam evaluations or segmenting the lungs so that the models learn only relevant features. The models described below this point, however, do not showcase any effort of such kind. Hence, even though the accuracies and other metrics of the models below this point might seem significantly higher than those mentioned above, the reader should keep in mind that they are not proven to generalize well in the real world.

To overcome the problem of a significantly smaller number of Covid-19 images as compared to Normal and CAP images, Sakib et al. [35] used a custom Generative Adversarial Network to generate more Covid-19 images for training. The dataset used by them consisted of 27228 Normal, 5794 CAP, and 209 Covid-19 images. On analysis, they found that generating precisely 100%, i.e., 209 new covid-19 images by GAN, led to the highest classification accuracy. On top of GAN, they used a customized CNN with ELU activation (Exponential Linear Unit) and Adagrad optimizer. The idea of using a customized and lean CNN works well in cases where data used for training is less. In such cases, even if the metrics are not necessarily excellent, we can be assured that the model will not overfit our small dataset, ensuring good generalizability. Talking about the results, this model achieved 93.94%, 88.52%, and 95.91% accuracy on Covid Positive, Community-acquired Pneumonia, and Normal cases, respectively.

Ali et al. [36] proposed a dual attention module to classify Viral pneumonia and Bacterial Pneumonia. For training, they used the popular dataset available on Kaggle, which consists of 5856 chest X-rays. The dual attention module consists of a spatial attention module and a channel attention module. For readers that do not know what "attention" is, attention was primarily used for NLP in Recurrent Neural Networks to allow the network to remember the relevant parts of a sentence. Later on, it was adopted into computer vision to determine the relevance of each feature with respect to the output. After that, each feature is multiplied by its weight to give importance to those features that contribute more to the output. The Channel Attention Modules measure the importance of each channel w.r.t other channels, whereas the spatial attention module measures the importance of each feature in a channel w.r.t other features in the same channel. This model achieved an accuracy of 97.82%.

Ohata et al. [37] used MobileNet to classify chest X-rays with covid-19 and normal chest X-rays. The dataset used consisted of 194 covid-19 images, and the normal images were collected from Kaggle and NIH datasets. They used MobileNet for feature extraction and tried six different classifiers for classification purposes. In the end, they decided to use Linear SVM for classification purposes which gave an accuracy of 98.62%. Lastly, Chowdhary et al. [38] tried using various models like SqueezeNet, MobileNet, InceptionV3, ResNet18, ResNet101, CheXNet, DenseNet201 and VGG19 on 423 Covid-19 images, 1579 Normal images and 1485 CAP images. They concluded that DenseNet and CheXNet perform best (99.70% accuracy) in two-class classification, i.e., Covid-19 and Other, whereas DenseNet performs best (97.94% accuracy) in three-class classification problems, i.e., Covid-19, CAP, and Normal. The sensitivity and specificity were 0.979 and 0.988, respectively.

| Author | Model | Task | Dataset | Accuracy |
|---|---|---|---|---|
| Haghanifar et al. [26] | U-Net + DenseNet121 | CP/N/CAP | 780/4600/5000 | 81.06% |
| Gu et al. [27] | FCN + (DCNN) | Bacterial/Viral | 2655/1848 | 76.92% |
| Zhang et al. [11] | ResNet + AD + CoP | CP/N/CAP | 5977/18619/18774 | 80.33% |
| Wang et al. [31] | POL + ResNet | CP/N/CAP | 204/2004/1314 | 99%/90%/93% |
| Arias-Londoño et al. [33] | U-Net + Covid-Net | CP/N/CAP | 8573/400/49000 | 91.53% |
| Sakib et al. [35] | GAN + Custom CNN | CP/N/CAP | 209/5794/27228 | 94%/88.5%/96% |
| Ali et al. [36] | ResNet + Attention | Bacterial/Viral | Kaggle | 97.82% |
| Ohata et al. [37] | MobileNet | CP/CN | 194/NIH-RSNA | 97.00% |
| Chowdhury et al. [38] | Multiple | CP/N/CAP | 423/1485/1579 | 97.94% |

*Table 3*

### 3.3 Localization of Pneumonia in Chest X-rays

While we have already covered some research that localized the entire lung region with the help of segmentation models like *U-Net* or *a-YoLo-like-lung-regressor*, it is worth noting that the research covered previously only localized the entire lung regions and not pneumonia-affected regions. Localization of pneumonia-affected regions in a chest X-ray can be beneficial in two ways. Mainly, it can assist radiologists in giving a quicker and more accurate diagnosis. Not only that, but localization also solves a significant problem of generalizability that we have encountered so far. If the primary goal of our DL model is to localize pneumonia-affected regions, we can be assured that the model is not looking at the wrong features to arrive at the right decision. As far as datasets are concerned, only one dataset (RSNA) has enough images with bounding boxes to train a DL that localizes well. Thus, it will be easy to compare all research work in this section based on metrics alone.

We start by explaining the approach of [39] because they won the RSNA Pneumonia Detection Challenge hosted by Kaggle. The authors used an ensemble of five models to localize pneumonia in chest X-rays. These five models were divided into two groups. The output regions from the first group (three models) were ensembled into one region. Similarly, the output regions from the second group (two models) were separately ensembled into a single region. Finally, the output regions from the two groups are ensembled into one output region using appropriate thresholds. The first group is made up of one Deformable Object Relation Network and two Deformable R-FCNs. Here, the prefix *Deformable* simply suggests the use of Deformable convolutions in the respective architectures. Deformable convolutions are different from regular convolutions in that every pixel/feature is offset by a certain amount in a certain direction. In this way, the shape of the receptive field of the convolution becomes free and is not limited to a rectangle. The offsets are learnable and thus play an essential role in correctly locating the entire object.

The Object Relation Network is not used very commonly and thus deserves some explanation. The Object Relation module is an adapted version of a basic attention module used in NLP. While the primitive elements of an NLP attention module are words, the primitive elements of an object relation module are objects. Because objects have a two-dimensional spatial arrangement and vary in terms of scale/shape, their locations and geometrical features are much more complex than the positions of words in a single sentence. Hence, the object relation module has an added geometric weight other than the original weight commonly found in NLP attention modules. The geometric weight considers the relative geometry of objects and models spatial relationships between them.

The second type of module used in the first group is Deformable R-FCN which is just R-FCN with Deformable convolutions. R-FCN is explained during the discussion of GeminiNet in this section itself.

Moving on, the second group is made up of two RetinaNets. The difference between these two RetinaNets is not in their architectures but in the type of input images used for training. The first RetinaNet, also called the ConcatRetinaNet, uses concatenated images for training. Each concatenated image is made by concatenating a pneumonia negative image with a pneumonia positive one. This way, the RetinaNet improves its distinguishing capacity while distinguishing between *lung opacity with pneumonia* and *lung opacity without pneumonia*. Images of 10 different sizes are given as input to all five models. Hierarchical ensembles are then formed from the two main groups, and finally, the bounding boxes from both models are ensembled according to different thresholds.

Li et al. [40] used 30000 images to train their model, and the rest of the images from the RSNA dataset were used for testing. Before using the raw images for input, they segmented the lung region from the original image using U-Net, much like [26]. After segmenting the lung region, they combined the segmented and raw images to make a final dataset for training their model. They used the SE-ResNet34 [41] for localizing regions containing pneumonia. SE-ResNet is short for Squeeze-and-Excitation ResNet, which is basically an encoder-decoder model that serves multiple purposes. The SE-ResNet acts as a feature extractor, and its side branch can automatically learn weights to assign importance to each channel. Moreover, the model can learn smoothly even over significantly deep layers without risk of degradation because of the residual blocks. Hence the model works as a Channel Attention Module over a ResNet34. For the final output, each pixel in the output channel represents the probability of that pixel belonging to the pneumonia class. The regions can then be extracted by applying thresholds to those probabilities. Coming to the results of this model, it was able to achieve an mAP score of 0.262. The mAP was calculated under IoU thresholds of 0.3, 0.4, 0.5, 0.6, 0.7.

Dimitrov Poplavskiy's team placed 2$^{nd}$ in the RSNA pneumonia detection challenge hosted by Kaggle. The paper written by [42] (including Poplavskiy) describes their model and approach in detail. For their model, they used RetinaNet, which is a single-shot detector. For the base of RetinaNet, they decided to use the encoder part of SE-ResNext-101. This particular design was chosen to accommodate both the speed of a single shot detector and the accuracy of a deep model like ResNext-101. Using this approach, they were able to achieve an mAP score of 0.26097. The official score on the leaderboard was 0.24781, but they optimized the model with heavy augmentations and zero rotation after the competition was over. A lot more trial and error went into making this model, mainly because it was made as a part of a competition. Almost all hyperparameters in this model are optimized and with good reasons, which are provided in their paper.

Up until now, we have talked about research that uses single-shot detectors for the localization of pneumonia-affected regions. However, two-stage detectors have a significant advantage over single-shot detectors in terms of accuracy. There is, of course, a time tradeoff involved while using two-stage detectors, but the question to ask is: how much does the detection time matter? At testing time, the difference between single-shot and two-stage detectors is not big enough to make any significant difference because real-time detection is not required for any use case of pneumonia localization.

Keeping this in mind, Yao et al. [43] presented the GeminiNet in March 2020. Before we begin with the explanation of this research work, there is a note worth taking. Some terminology in the following four or five sentences might sound new to beginners, but all of it is elaborated upon in considerable detail in the two successive paragraphs. Continuing with GeminiNet, it is a two-stage detector that builds upon the concept of R-FCN [44]. The difference between R-FCN and GeminiNet is that the latter uses RFB [45] blocks instead of simple convolution blocks for multi-scale context information. Moreover, they changed the base model used for feature extraction. Instead of using ResNet-50, they used DetNet59 because it yielded better performance metrics. This model (DetNet59 + GeminiNet) presented by the authors achieved an mAP score of 0.3259 at IOU thresholds 0.4,0.5,0.6 and 0.7.

Now onto the elaboration, the RFB block is much like an InceptionV1 block, except it has an extra shortcut like residual blocks in a ResNet. RFB blocks are especially useful in object detection scenarios because they have variable receptive fields (like Inception), and they can handle deep models smoothly (like ResNet). Moreover, instead of simple convolutions, the authors used dilated convolutions [46] in the RFB block. Dilated convolutions convolve upon a larger size (say 5x5 instead of 3x3) but select only a few features (3x3 = 9) from the big block (5x5), thereby keeping the number of parameters small but increasing the receptive field.

While RFB blocks are important in GeminiNet, its heart is the R-FCN. R-FCN is short for Region-Based Fully Convolution Network, and it is used as a substitute for Fast-RCNN and Faster R-CNN. Fast R-CNN improves upon the speed of R-CNN by calculating the feature map of the entire image at once and uses that feature map to derive ROI's directly. This way, feature maps do not have to be calculated for different ROIs separately. R-FCN works by simultaneously generating ROI's and region-based feature maps, thus saving a lot of time. After that step, for all regions generated in the ROI step, region-based feature maps are checked to vote for the probability of a particular ROI containing a particular part of the entire object. The final vote array (Consisting of probabilities from all ROI's) is averaged to determine which object is present in the image. This process of calculating probabilities for all ROI's and storing them in a vote array is called PS-ROI (Position Sensitive ROI) pooling. GeminiNet does not use R-FCN as it is. The changes are as shown in figure 2.

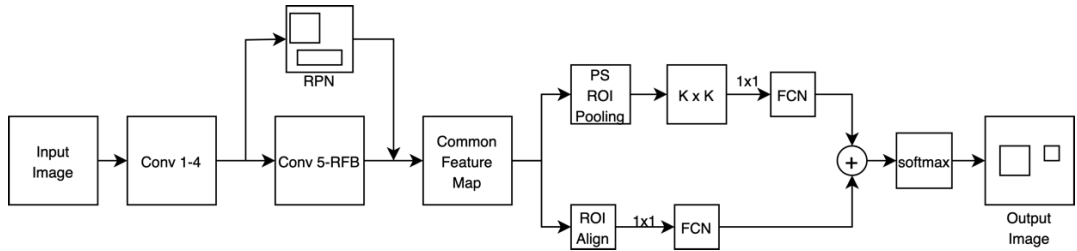

*Figure 2*

While on the topic of R-FCN, the approach of [47] is worth mentioning. They used a modified version of R-FCN called CoupleNet [48]. CoupleNet adds a second branch to R-FCN for processing global features. This way, the resulting architecture learns features from a larger area through the global branch by adding extra ROI features, and local features learn from the local branch by using PS-ROI features. The DeepRadiology Team used an ensemble of four models having the same architecture. All four of these models gave unique outputs which were used for generating the final regions. First, all bounding boxes that had a confidence score less than 0.5 were eliminated. After that, bounding boxes from all four groups which had IOU greater than 0.25 were grouped together. Lastly, the coordinates of all bounding boxes in one group were used to derive a final bounding box. This model was able to achieve an mAP of 0.23089 and placed 7$^{th}$ in the competition.

Next, we move on to models that use a combination of Single Shot Detector and Two-Stage detectors. Sirazitdinov et al. [49] presented a model that used a combination of RetinaNet (Single-Shot detector) and Mask R-CNN (Two-Stage detector). RetinaNet worked as the main unit, while Mask R-CNN was used as an auxiliary unit to adjust the regions of RetinaNet. The working of the entire model is straightforward. Both the RetinaNet and the Mask R-CNN models work separately and predict bounding boxes with corresponding classes. After applying Non-Max Suppression in both models, a weighted average of predictions from both models is calculated where the weight of RetinaNet: Mask R-CNN predictions is 3:1. This ratio was calculated by an iterative grid search over many such ratios ranging from 1:1 to 4:1.

Another research work exploring the combination of RetinaNet and Mask R-CNN for pneumonia detection is due to [50]. They tried various ensembles of RetinaNet and Mask R-CNN with different sizes and different weights. Finally, a model with RetinaNet 178, RetinaNet 184, RetinaNet 201, Mask R-CNN 150, and Mask R-CNN 162 in the ratio 2:2:3:2:3 was used for detection. This model achieved an mAP of 0.21746, which could be placed at the 21$^{st}$ place in the competition approximately.

| Author | Model | Type | IOU Thresholds | mAP |
|---|---|---|---|---|
| Li et al. [40] | U-Net (SE-ResNet34) | SSD | 0.3-0.7 (0.1) | 0.262 |
| Gabruseva et al. [42] | RetinaNet (SE-ResNext101) | SSD | 0.4-0.75 (0.05) | 0.260 |
| Yao et al. [43] | GeminiNet (modified R-FCN) | TSD | 0.4-0.7 (0.1) | 0.326 |
| The DeepRadiology Team [47] | CoupleNet (modified R-FCN) | TSD | 0.4-0.75 (0.05) | 0.231 |
| Sirazitdinov et al. [49] | RetinaNet + Mask R-CNN (3:1) | SSD + TSD | 0.4-0.75 (0.05) | 0.204 |
| Ko et al. [50] | RetinaNet + Mask R-CNN (7:5) | SSD + TSD | 0.4-0.75 (0.05) | 0.217 |
| Pan et al. [39] | R-FCN + RelNet + RetinaNet | SSD + TSD | 0.4-0.75 (0.05) | 0.255 |

*Table 4*

## 4.1 Classification of Covid-19 and CAP via CT scans

Harmon et al. [51] made a Deep Learning model detect covid-19 from CT scans using multinational datasets. Their dataset consisted of Covid Positive scans from China (369), Japan (100), and Italy (57). In total, 1059 scans were used for training, and 1397 separate scans were used for testing. Their DL model consists of a lung segmentation module and a classifier module. The lung segmentation module segments the lung region from the entire CT scan. After the lung region is segmented, the segmented region is given as an input to the classifier, which classifies the input into Covid Positive or Covid Negative. For the lung segmentation module, the AH-net [52] architecture is used. AH-Net is an encoder-decoder based segmentation module used for 3D segmentation, and it mostly works like U-Net. The segmented regions used while training had a mean dice score of 0.95. Dice scores are similar to IOU scores and are used widely as a metric in segmentation tasks. Moving on, the classification module is made up of the DenseNet 121 architecture just like CheXNet and takes a fixed input of size 192 x 192 x 64. Finally, this model achieved an accuracy of 89.6% with an AUC score of 0.941 on independent testing sets. While the architecture of the classifier in this model is the same as CheXNet, the number of training images is significantly fewer. Nevertheless, Grad-CAM evaluations of this model show that the model can learn correct features to arrive at the right decision. Hence, the segmentation module that precedes the classification module plays a vital role in the generalizability of this model. The sensitivity and specificity of this model were 0.840 and 0.930, respectively.

Ouyang et al. [53] presented a DL model with dual sampling and an online, trainable CAM module to ensure that the model learned important features. The training dataset used for this model contains 2186 images, of which 1092 are Covid Positive, and 1094 are CAP. The dataset used for testing is also quite large, with 2796 images, of which 2295 are Covid Positive, and 501 are CAP. The authors also use a standard lung segmentation module called the VB-Net toolkit [54] for lung segmentation. Feature extraction is then done using a ResNet34. After segmentation, the entire dataset is sampled in two ways. The first one is uniform sampling, where each minibatch contains images in the same ratio as the entire dataset. The second method is size-balanced sampling. Size-balanced sampling is required because the dataset has only a small number of Covid-19 images with a small infection area. Similarly, only a few images with a large area of infections are available in the CAP category. Hence, size-balanced sampling is applied such that the ratio of 1) CAP images with large infection 2) CAP images with small infection 3) Covid images with large infection 4) Covid images with small infection remain approximately the same in each minibatch. This ratio is maintained by over-sampling. However, oversampling poses another challenge of overfitting. This challenge is resolved by using the first of its kind, online CAM (class activation mapping) module. The online CAM module is generated by applying a 1x1x1 convolution to the weights of the fully connected layer and then convolving that layer over the feature map. A ReLU operation is applied at last to get the final activation map. This model achieved 95.4% accuracy with an AUC of 0.988. The sensitivity and specificity of this model were 0.872 and 0.907, respectively.

The work of [55] is yet another example of a DL model that consists of a lung segmentation module followed by a classifier with attention. Their dataset consists of 4657 scans where 936 are Normal, 2406 are CAP, and 1315 are Covid Positive. For segmentation, the authors used the 3D-UNet [56] models. After lung lobe segmentation, the images are cropped into a size of 96x96x96 and passed into the classifier. The classifier consists of two parts, the pneumonia detector and the pneumonia classifier. If an image is detected to have pneumonia by the pneumonia detector, it is passed to the pneumonia classifier, which classifies the image into ILD (Interstitial Lung Disease) or Covid-19. The fact that the pneumonia classifier only comes into action after the pneumonia detector has performed its job was leveraged into using a Prior Attention Residual Block. As shown in figure 3, the prior attention residual block has one additional input other than the regular residual block, which is borrowed from the weights of the final layer of the pneumonia detection module. This way, the prior attention residual block can get the attention weights before backpropagation takes place, and they can be used to train the pneumonia classifier simultaneously. This method ensures that the classifier is trained on the right features. This model achieved an accuracy of 93.3% on the Covid-19 class, 89.4% on the ILD class, and 91.5% on the Normal Class. The sensitivity and specificity for Normal/Viral/Covid-19 classes were (91.5/89.4/93.3) and (93.5/90.6/95.5), respectively.

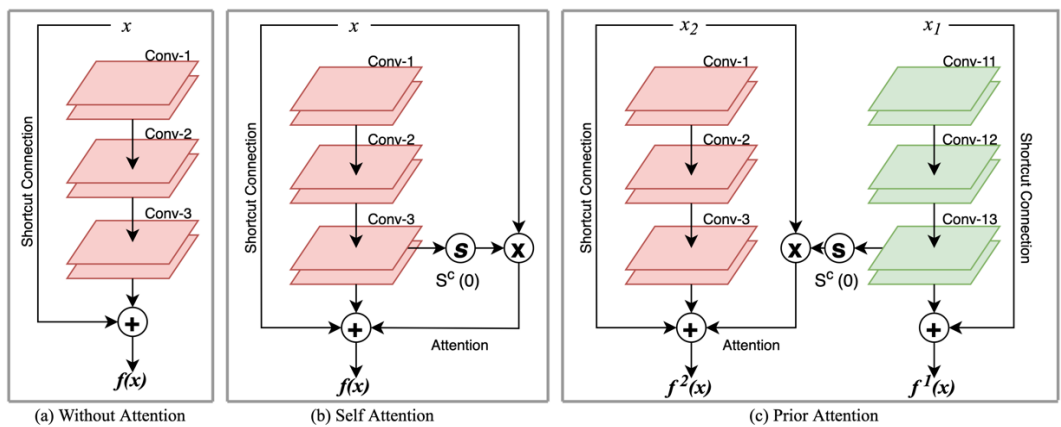

*Figure 3*

Lai et al. [57] proposed the NCIP-Net for the detection of Covid-19 from CT scans. Before we move on, NCIP is short for Novel Coronavirus-Infected Pneumonia. The authors of NCIP-Net used a multi-task Deep Convolution Neural Network for 1) determining the presence of Covid-19 based on the entire image, 2) Segmentation of Covid-19 lesions from the entire CT scan, and 3) determining the probability of Covid-19 from the segmented lesions. The dataset used for training this model consists of 323 Covid-19 positive CT scans and 501 Normal scans. Before providing the images to the model as an input, all images went through a lung lobe segmentation process where the lung region was separated from the entire image. The model is constructed like a normal encoder-decoder, but the encoder is connected to three branches. Out of those three branches, one is the decoder which is used for lesion segmentation. The second branch from the encoder is used for the prediction of Covid-19 directly from the image. The third branch is used for determining the probability of Covid-19 based on the ROI with lesions. The training is divided into two stages. In the first stage, the second branch from the encoder is connected to three convolution layers with a residual block concatenated with a softmax function to determine the probability of Covid-19 from the image directly. Still, in the first training stage, the features encoded by the encoder are passed on to the decoder for lesion segmentation based on dice loss. In the second stage of training, CT volume patches are used as an input and the third branch extended from the encoder (C-Net) is used to identify a maximum of 10 proposals with the likelihood of lesions to predict the presence of Covid-19. The encoder can predict the proposals with the likelihood of lesions because it was previously trained to segment lesions from the CT scan. This model achieved an accuracy of 74.4% in Covid-19/Normal and 82.9% in Covid-19/Other Lung Diseases.

Looking at all this research work, some patterns clearly stand out. The first and the most important one is to segment the lung region from the entire CT scan. This way, a lot of computation time is saved, and the model is forced to learn features from the right region. However, the model can still learn the wrong features from the lung region. To overcome this problem, some kind of attention mechanism, online or offline, is used in all models that are proven to generalize well. Next, we move on to some research work that distinguishes pneumonia from normal cases and does not include covid-19 cases. A separate section was not created to include the detection of pneumonia via CT scans because not enough research has been carried on that topic. This is because detection of pneumonia is usually done with X-rays rather than with CT scans.

Wang et al. [58] proposed a multi-channel multi-modal deep regression framework for the screening of pneumonia from CT scans. For their model, they used 450 pneumonia-positive CT scans and 450 Normal CT scans. Not only that, but they also used the complaints of those patients and their demographic information to improve the performance of their model. The entire model is divided into three parts that process demographic information, complaint information, and CT scans, respectively. Intuitively, the demographic information and the complaint information are processed with the help of an LSTM. The CT scans, however, are processed differently. First, three slices from the CT, namely the Lung Window (LW), High Attenuation (HA), and Low Attenuation (LA), are extracted and concatenated into a three-channel image. This three-channel image is then passed onto an RCNN with a base of ResNet-50. The RCNN is an object detection module, so it detects the region of the CT scan where pneumonia is present. The features extracted from the region detected by the RCNN are then passed on to an LSTM network. The features extracted by the RCNN were passed on to the LSTM for two reasons. First, the authors wanted to use the three channels as a sequence of video frames that were dependent on each other. The second reason is that an LSTM was the only feasible way to concatenate the demographic and complaint information with the spatial information of CT scans. Finally, all three LSTMs are concatenated and used for pneumonia detection. This model achieved an accuracy of 94.6% in the pneumonia detection task. The sensitivity and specificity of this model were 0.933 and 0.922, respectively.

| Author | Model | Task | Dataset | Accuracy |
|---|---|---|---|---|
| Harmon et al. [51] | AH-Net + CheXNet | CP/N/CAP | 1059 | 89.6% |
| Ouyang et al. [53] | VB-Net + ResNet34 | CP/CAP | 1092/1094 | 95.4% |
| Wang et al. [55] | 3D U-Net + ResNet | CP/N/CAP | 1315/936/2406 | 93.3%/91.5%/89.4% |
| Lai et al. [57] | NCIP-Net | CP/N | 323/501 | 74.4% |
| Wang et al. [58] | ResNet + LSTM | N/CAP | 450/450 | 94.6% |

*Table 5*

## 4.2 Localization of Covid-19 in CT scans

Wang et al. [59] presented the COPLE-Net, a noise-robust model for segmentation of covid-19 lesions from CT images. To train their model, they used 558 Covid Positive CT images. The architecture of COPLE-Net was based on U-Net with some modifications. First, instead of using only max-pooling or average-pooling for downsampling, the authors concatenated both methods, and it gave better results. Second, they modified the skip connections of U-Net by adding another layer of convolution between the encoder and the decoder. This additional layer contains half as many channels as the encoder. This layer was added to alleviate the semantic gap between the decoder's high-level features and the encoder's low-level features by forcing the encoder features to a lower dimension (half channels). Third, the authors added an Atrous Spatial Pyramid Pooling (ASPP) [46] layer at the end of the encoder. An ASPP layer contains four parallel layers of dilated convolutions with different dilation rates. This way, multi-scale features can be extracted for small and large lesion segmentation.

COPLE-Net was trained using an adaptive self-ensembling technique with a noise-robust dice loss. The noise-robustness in dice loss was achieved by using an MAE analogous dice loss instead of the usual MSE analogous dice loss. To understand the self-ensembling, we must first understand which models were ensembled. The authors trained two COPLE-Nets via a teacher-student mechanism. The teacher model was an exponential moving average of the student model and was thus more stable than the student model. However, the weights of the moving average were not fixed from the beginning. If the loss of the student model was more than a defined threshold, the student model was not used to update the teacher model at all. Otherwise, the weight of the student model considered to update the teacher model was defined as a function of the loss constant (difference between the losses) of the said models. This model was able to achieve a dice score of 0.8072 or 80.72%.

Gao et al. [60] presented a Dual-branch Combination Network (DCN) for performing lesion segmentation and classification at once. Their dataset consisted of 1918 CT scans from 1202 subjects across 2 hospitals. Before feeding the CT image slices into the DCN, the images underwent lung segmentation through a U-Net. These segmented lungs with a dice score coefficient of 0.99 were then used as an input to the DCN model. The model comprises two main parts, one for classification and another one for segmentation. The segmentation model is an encoder-decoder model analogous to a U-Net model. The classification model uses ResNet-50 as a backbone with Lesion Attention modules, as shown by brown color in figure 4. The LA module is a combination of 1) (the original CT slice)/(ResNet-50 down-sampled slice) and 2) the feature extracted slice of the corresponding size from the decoder of the segmentation module. A slice from the decoder module is chosen because the decoder has more relevant features which correspond to covid-19 lesions. Hence, the ResNet-50 classification module is forced to pay attention to features that contain covid-19 lesions. This model was able to achieve a dice score of 0.8351 or 83.51%. The classification accuracy for internal validation (CT images from the same hospital that the model was trained on) was 96.74%, with an AUC of 0.9864, while the accuracy on external validation (CT images from a different hospital) was 92.87% with an AUC of 0.9771.

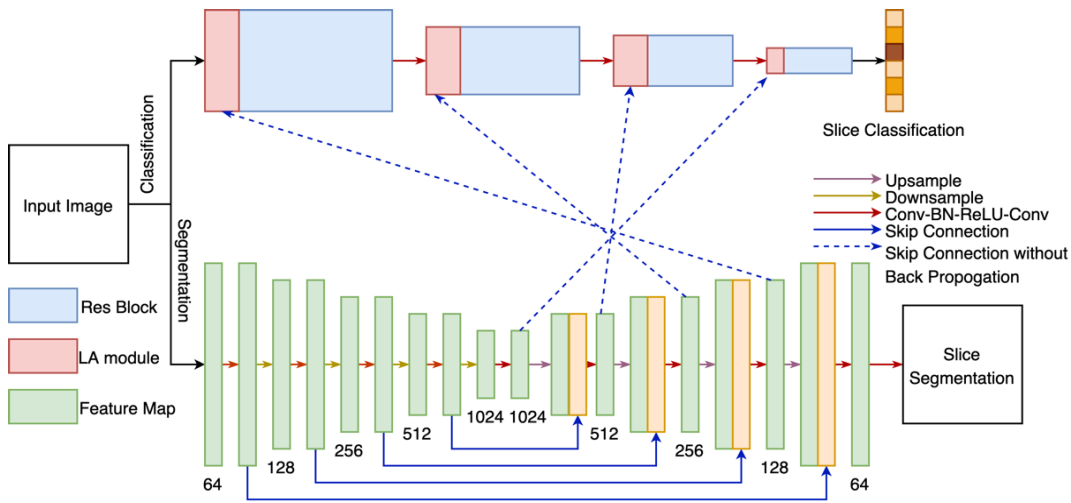

*Figure 4*

Zhou et al. [61] presented a three-way segmentation technique for segmentation of covid-19 infected regions from a CT scan. The dataset used by them consisted of CT scans of 120 patients. The total number of unique CT scans used is not disclosed in their paper. The authors, however, used a unique data augmentation technique to generate 200 CT scans from each unique patient. The detailed augmentation technique has not been disclosed in the paper, but the principles upon which the augmentation was based were delineated. Hence, the dataset consists of approximately 24000 CT scans. The authors used three-way segmentation in that they extracted x-y, y-z, and x-z slices from the CT scan and trained three different segmentation models to segment covid-19 lesions from these models. This technique is analogous to how radiologists diagnose covid-19 lesions. If a particular voxel cannot be clearly predicted as lesion or normal, radiologists often look at voxels surrounding that voxel. Similarly, if we have 2D segmentations from all three axes (x-y, y-z & x-z), our model can classify a voxel into lesion or normal by looking at surrounding voxels without being limited to that particular plane. This model was able to achieve a dice score of 0.783.

Fan et al. [62] presented the Inf-Net, a semi-supervised deep learning model for the segmentation of Covid-19 lesions from CT scans. Their dataset consisted of 50 CT scans which aptly justifies the semi-supervised learning. The architecture of Inf-Net begins with two convolution layers into which a CT scan slice is fed. The first two convolution layers extract the low-level features. Generally, low-level features are known to detect edges in computer vision, so these features are passed through a simple convolution layer and compared against the ground truth segmented region to determine the edge loss. As shown in figure 5, this edge loss is backpropagated to f2 so that f2 can learn correct edge features. Next, the features of convolution layers 3,4, and 5 are passed on to a partial decoder which yields a coarse global map of the region to be segmented. Only high-level features are used as an input to the partial decoder because [63] pointed out that low-level features are computationally intensive as compared to high-level features and contribute little to the process of segmentation. The global map provided by the partial decoder is labeled as coarse in that it contains an extra segmentation region that needs to be removed. Hence, a reverse attention module is used to erase the extra region from the coarse global map. The removal of this extra region is done with the help of edge features from the second convolution layer so that only the region inside the edge is preserved. Therefore, the reverse attention module takes input from both f2 and the global coarse map. Three such reverse attention modules, R3, R4, and R5, are stacked in a cascade manner such that the output of R5 is used as an input for the reverse attention module of R4 and so on. Finally, the output of R3 is followed by a sigmoid function to give the completely segmented infected region. The semi-supervised learning approach of Inf-net is progressively enlarging the dataset. This process is performed by predicting some labels from the limited training data and then using the predicted labels as the training data and the original training data. This process is repeated for a while until enough training data is gathered. The Inf-Net achieved a dice score of 0.739 on their dataset and a dice score of 0.597 on a different dataset.

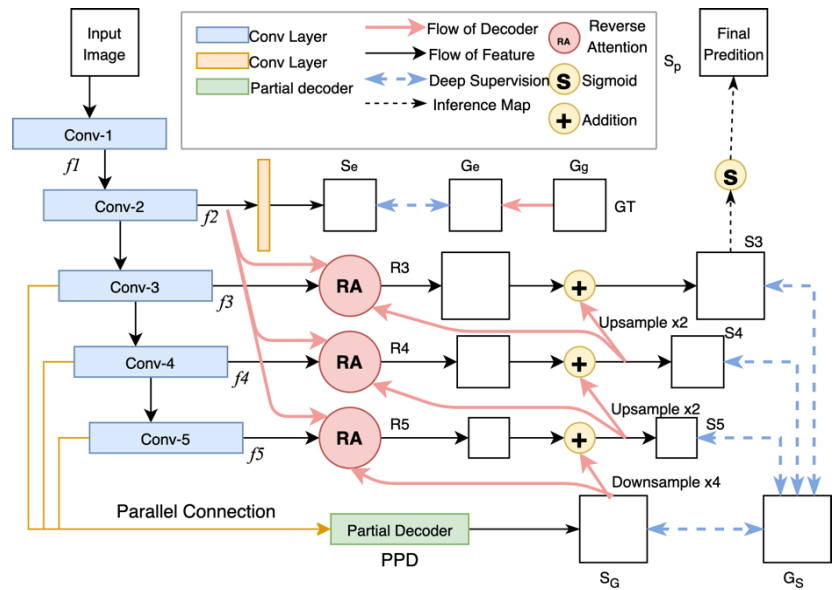

*Figure 5*

Yang et al. [64] presented a unique approach for the localization of Covid-19 lesions in CT slices. The idea was to train a Generator Network, which would output Normal (without covid-19) slices even if the corresponding input slice had covid-19 lesions. Afterward, the output slices could be subtracted from the input slices to localize the regions where covid-19 lesions were present. The generator model was trained against a discriminator model, which tried to distinguish between real and generated normal pneumonia images. Moreover, a ResNet-18 was also trained on covid-19 positive images so that the ResNet could grasp the low-level features and concatenate those features with the encoder of the generator network. This was done because the Generator network itself was not powerful enough to grasp the low-level features of a CT slice. Finally, both normal and covid-19 positive CT slices are provided to the generator model, but the loss is only calculated against normal images. In this way, the generator is forced to generate normal CT slices even from the covid-19 lesion containing CT slices. This is analogous to a denoising autoencoder where noisy images are passed into the auto-encoder, but the loss is calculated against noise-less images. A major benefit of using this model is that it is weakly supervised. Hence, while training the generator, labeled image pairs are not necessarily required. This model achieved a dice score of 0.575, which is very competitive for weakly supervised models. However, fully supervised models have a much higher dice score.

| Author | Model | Type | Dataset | DSC |
| --- | --- | --- | --- | --- |
| Wang et al. [59] | COPLE-Net (Modified U-Net + ASE) | Fully Supervised | 558 scans | 0.8072 |
| Gao et al. [60] | DCN (Modified U-Net + LA + ResNet) | Fully Supervised | 1918 scans | 0.8351 |
| Zhou et al. [61] | U-Net (X-Y, Y-Z, X-Z axes segmentation) | Semi-Supervised | 120 patients | 0.783 |
| Fan et al. [62] | Inf-Net (Custom CNN + RA + PD) | Semi-Supervised | 50 scans | 0.594 |
| Yang et al. [64] | GAN + ResNet | Semi-Supervised | 1252 scans | 0.575 |

*DSC = Dice Score Coefficient; ASE = Adaptive Self Ensembling; LA = Lesion Attention; RA = Reverse Attention; PD = Partial Decoder*

Table 6

## 3. Challenges and Future Scope

The end goal of all research into automatic pneumonia/covid-19 detection and localization is to have a model that can be used in (hospitals)/ (chest X-ray centers)/ (CT scan centers) on an everyday basis. For a single model to be used in different centers worldwide, the model should be able to generalize well to different CT scan/X-ray machines and different demographics.

This poses the problem of collecting a dataset that contains such a wide variety of data. While the problem of overfitting to a particular dataset has been mitigated by attention mechanisms, Grad-CAM analysis, adversarial training, and segmentation-before-classification, this kind of work needs to be applied to a more distributed dataset so that it can learn correct features from any chest X-ray/CT scan around the world without the need of tedious pre-processing. Hence, the first future scope would be to collect a dataset with a wide variety of chest X-rays/CT scans, especially for covid-19 classification.

Pre-processing an image of a chest-X-ray/CT-scan before using it as an input for a DL model poses another challenge because most image pre-processing is dependent on the type of image. For example, chest X-rays taken on machine A would require a different kind of image pre-processing mechanism than a chest X-ray taken on machine B. Hence, another future scope would be creating DL models which require little to no data-dependent pre-processing.

In this work, a lot of different research that tackles different problems has been illustrated. Although no single work tackles all challenges, a smart combination of some practices used in the mentioned research might yield a truly generalizable model. Furthermore, several small, custom datasets were compiled by different authors for their research. Combining those datasets or even using semi-supervised domain adversarial training with different datasets would generalize the corresponding DL model better.

Practical application of research in such DL models might be restricted to assisting doctors in making a better diagnosis instead of working in complete autonomy. Keeping such applications in mind, DL models can be modified to output a prediction highlighting the most important features based on which the prediction was made. This way, doctors might get help if they miss some features in the image which are not apparent to the naked eye.

## 4. Conclusion

The process for automating the detection of pneumonia from chest X-rays and CT scans has evolved a lot over the past few years, especially with the advent of Deep Learning methods. Looking back at the past four years, base DL model architectures have evolved a lot. However, base model architectures are not the most effective solutions for the specific task of pneumonia detection. The pioneering models that achieved good metrics on pneumonia detection tasks tweaked the architectures of base models so that the tweaked models were a better fit for the task of pneumonia detection. The models that followed these pioneering models were focused on generalizing the model architecture. This generalization was achieved through techniques like adversarial training, Grad-CAM analysis, attention mechanisms, and many more.

The task of classifying covid-19 from chest X-rays and CT scans is not very different from the pneumonia detection task. However, research into covid-19 detection through DL models is relatively new because covid-19 is a relatively new disease (as of 2021). Because of the time gap, the models made for detecting covid-19 from pneumonia use better base model architectures than those initially used in pneumonia detection. However, the techniques used to make the base models more effective toward the specific task of covid-19 detection are similar to the techniques used for the pneumonia detection task, both for higher metrics and better generalization. This observation leads us to an important inference. The inference would be that those techniques which make base model architectures more effective or more generalizable for a specific task (pneumonia detection) are at least as important if not more important than the base models.

Even as base model architectures keep improving, the techniques discussed in this paper can always be applied to the improved base models to further improve the base models' generalizability and effectiveness. With that thought, many different techniques and architecture tweaks, along with their merits, demerits, and tradeoffs, have been explained in this paper. A quantitative analysis table that corresponds to each subsection of the papers is also provided so that the readers can co-relate between the qualitative and quantitative results of different models and techniques. With both qualitative and quantitative analysis, this paper can be a one-stop solution for aspiring researchers who want to study the field of pneumonia/covid-19 detection in depth. Lastly, this paper serves as a means of initiating and propagating new research in the field of automatic pneumonia/covid-19 detection and localization by providing a wide breadth of techniques along with enough depth in every technique so as to guide aspiring researchers in the right direction for their specific purpose.